# Symmetry Breaking by Interfacial Dead Layers: Observation of Forbidden Self-Induced Spin-Orbit Torque in Symmetric Ferromagnets.

Athira Ravindran K*,[1] Raghvendra Posti*, †,[1] Manish Kumar Mohanta,[2,6] Abhishek Kumar,[1] Damanpreet Kaur,[3] Alberto Anadón,[4,5] Sebastien Petit-Watelot,[4] Juan-Carlos Rojas-Sánchez,[4] Michel Hehn,[4] Puru Jena,[2] and Debangsu Roy[1]

[1]*Department of Physics, Indian Institute of Technology Ropar, Rupnagar 140001, India*

[2]*Department of Physics, Virginia Commonwealth University, Richmond, Virginia 23284, USA*

[3]*Department of Physics and Materials Science and Engineering, Jaypee Institute of Information Technology Noida, Uttar Pradesh 201309, India*

[4]*Université de Lorraine, CNRS, IJL, Nancy, F-54000 France*

[5]*Instituto de Nanociencia y Materiales de Aragón (INMA), CSIC-Universidad de Zaragoza, Zarazoza 50009, Spain*

[6]*Department of Physics, Indian Institute of Technology Bhubaneswar, Bhubaneswar 752050, Odisha, India*

Conventionally Spin-orbit torques (SOTs) in ferromagnets require heavy-metal layers or engineered structural asymmetry to break inversion symmetry. In this work, we report the observation of robust, self-generated SOTs in a nominally symmetric, heavy-metal-free MgO/NiFe/MgO trilayer—a geometry where such torques are theoretically forbidden. By combining harmonic Hall measurements with SQUID magnetometry and X-ray photoelectron spectroscopy, we identify the symmetry-breaking origin: a ~1.8 nm magnetic dead layer at the bottom interface. Crucially, we demonstrate a quantitative agreement between our data and the drift-diffusion theory predicted by Kim and Lee[1], yielding a theoretically extracted dead-layer thickness (~1.2 nm) which matches structural characterization. Furthermore, density-functional calculations confirm that NiFe possesses sufficient intrinsic spin Hall conductivity to support the observed spin currents. These results reframe the "parasitic" dead layer as a functional spintronic component, establishing a universal, all-ferromagnetic route to SOTs in standard magnetic heterostructures.

*equal contribution

†*current affiliation: Department of Physics, Carnegie Mellon University, Pittsburgh, 15213, PA, United States*

## INTRODUCTION

Ferromagnets (FMs) constitute the foundation of spintronics, where the efficient manipulation of magnetization via spin-orbit torques (SOTs) remains a primary goal[2-6]. Generally, SOT generation relies on heavy metal (HM)/FM heterostructures[7-10], where the non-magnetic HM generates a transverse spin current via strong spin-orbit coupling (SOC). However, the high symmetry of these bilayers restricts the allowed spin polarizations to be orthogonal to both the charge current and the electric field[11]. In contrast, ferromagnets intrinsically break time-reversal symmetry[11-14], permitting unconventional spin polarizations[15-26] and potentially enabling "self-torques" generated within the FM itself[27-42], eliminating the need for heavy metals.

Despite this potential, a fundamental symmetry constraint remains: in a structurally symmetric ferromagnetic film, spin currents generated within the bulk accumulate equally and oppositely at the top and bottom interfaces, cancelling the net torque[11,12]. Thus, experimental realizations of self-torque have largely relied on deliberately breaking global inversion symmetry, either by engineering steep compositional gradients[28,30,43-45] or by utilizing exotic low-symmetry crystal structures.

A pivotal advance in this framework was the theoretical work of Kim and Lee[1], who predicted that a net self-torque could arise in a single FM layer solely through asymmetry in interfacial spin absorption rates, even in the absence of bulk gradients. This theory challenges the necessity of heavy metals or bulk asymmetry. However, to date, experimental validation has been limited due to the complexity of the materials studied. In order to corroborate the same a simple system with nominal structural symmetry and low intrinsic SOC is required, where the torque can be unambiguously linked to the interfacial asymmetry predicted by theory.

In this work, we have used a symmetric MgO/NiFe/MgO trilayer—a "standard model" system where net SOT generation is theoretically forbidden to validate the proposed theoretical model of Kim and Lee. Contrary to standard expectations, we observe a robust damping-like torque comparable to HM-based systems together with a non-negligible field-like torque, which is expected even in symmetric structures due to spin precession within the ferromagnet[1].We identify the origin of this torque as a naturally occurring magnetic "dead layer" at the bottom interface, which acts as the symmetry-breaking element.

Moreover, we substantiate this mechanism through the comparison of transport data with theoretical models. By fitting the thickness dependence of the observed SOT to the Kim-Lee drift-diffusion formalism, we extract a theoretical dead-layer thickness of ~1.2 nm. This value is in good agreement with the physical dead-layer thickness of ~1.8 nm independently measured via SQUID magnetometry and X-ray photoelectron spectroscopy (XPS). Furthermore, our density functional theory (DFT) calculations confirm that the NiFe lattice intrinsically supports the spin Hall conductivity required to drive these currents. This quantitative correspondence provides the direct experimental confirmation that a ubiquitous interfacial dead layer can create the specific asymmetry in spin-mixing conductance required by the Kim-Lee theory. Our work thus redefines "parasitic" dead layers as functional, intrinsic sources of symmetry breaking, establishing a universal route to SOTs in standard magnetic heterostructures.

**EXPERIMENTAL DETAILS**

We fabricated a MgO (2 nm)/NiFe (t nm)/MgO (3.5 nm) trilayer structure, where the thickness of the NiFe layer, t, was varied from 4.5 nm to 14.4 nm. The stack was deposited on a 3-inch thermally oxidized silicon wafer using magnetron sputtering. To achieve systematic thickness variation, we employed a wedge deposition technique for the NiFe layer by depositing in a 5 cm long substrate at an angle with respect to the target, preparing two sets of samples with gradient profiles of 0.59 nm/cm and 1.31 nm/cm, respectively. This approach enabled a detailed study of thickness-dependent effects across a single wafer while maintaining uniform growth conditions. The MgO layers on either side of the NiFe serve two key roles: they protect the NiFe from oxidation and maintain structural symmetry at the interfaces, minimizing artificial inversion symmetry breaking.

Six-terminal Hall bar devices [Fig. 1(a)] with a current channel width of 20 μm and a length of 100 μm were patterned using standard photolithography followed by plasma etching. In our design, current channels were fabricated both along and perpendicular to the NiFe thickness gradient direction. Notably, devices with current channels aligned parallel to the wedge (gradient) direction exhibited negligible thickness variation along the *y*-axis [Fig. 1(a)] and were thus considered effectively uniform in NiFe thickness.

Electrical contacts were fabricated using thermally evaporated Au pads. Micro-wire bonding was subsequently performed to enable room-temperature transport measurements. A schematic of the Hall bar geometry and measurement configuration is shown in Figure 1a along with the optical microscope image. The current was applied along the x-direction ($\hat{x}$), and transverse Hall voltages or resistances were measured across the lateral Hall terminals. All transport measurements were conducted at room temperature. For DC measurements, we used a Keithley 6221 current source and 2182A nanovoltmeter, and AC harmonic measurements were performed using 6221 in combination with an EG&G 7265 lock-in amplifier referenced at 577.13 Hz.

The magnetic anisotropy of the samples was characterized using a combination of magneto transport and magnetometry techniques. In-plane magnetic anisotropy (IMA) was confirmed by both superconducting quantum interference device (SQUID) magnetometry and Hall measurements. Figure 1b displays the in-plane magnetization (M-H) loop measured via SQUID for a representative sample with NiFe thickness t = 4.5 nm, indicating a square loop with high remanence. The inset shows the planar Hall resistance ($R_H$) as a function of in-plane rotation angle ($\phi$) under constant external magnetic field of 2000 Oe, exhibiting a characteristic sin2$\phi$ dependence consistent with in-plane anisotropy.

Out of plane magnetization behavior is shown in Fig.1(c) where the M-H loop confirms that the hard axis lies perpendicular to the film plane. The inset shows the anomalous Hall Resistance ($R_H$ ) as a function of out-of-plane magnetic field ($H_Z$), supporting the absence of perpendicular magnetic anisotropy. Minor deviations near zero field are attributed to domain effects or background subtraction artifacts, which do not influence the extracted high-field magnetic parameters. These combined results confirm that the magnetization preferentially lies in the film plane, and further details are provided in the supplementary information.

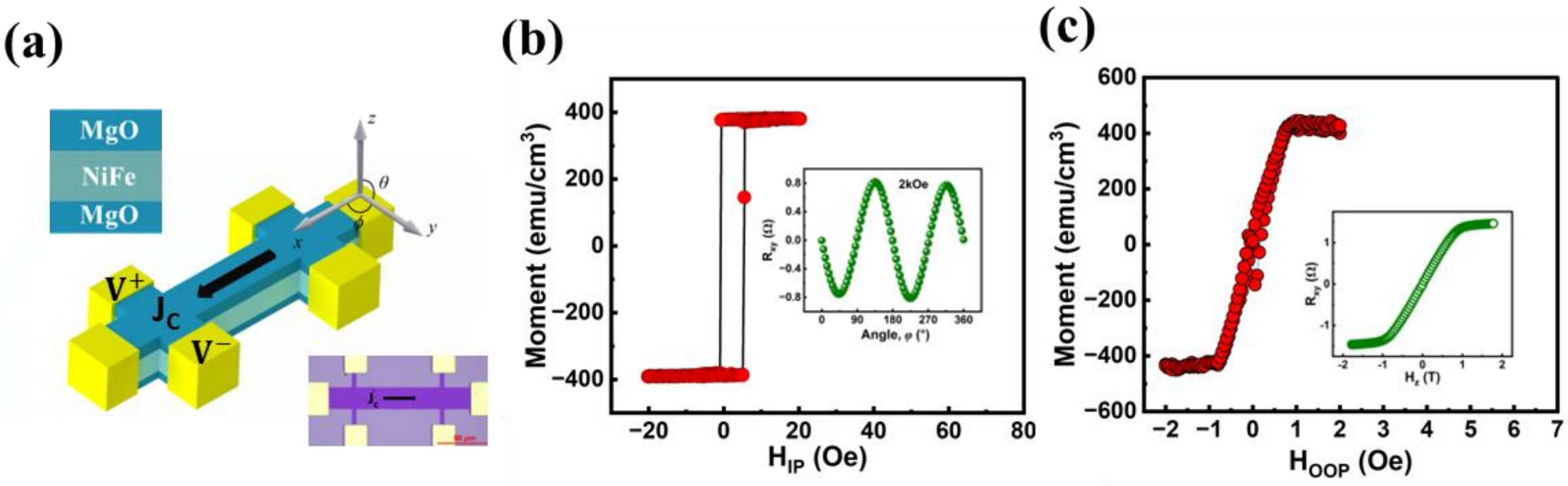

FIG. 1. (a) Schematic of the six-terminal Hall bar geometry used for the experiments with MgO/NiFe/MgO stack and an optical microscope image of the Hall bar used for SOT characterization. (b) In plane

magnetization curve for a 4.5nm thick NiFe film (inset: planar Hall voltage measured as a function of in plane rotation angle under a fixed external field of 2kOe), (c) Out of plane magnetization curve for the same film (inset: anomalous Hall voltage measured as a function of out of plane magnetic field).

Next, for systematic thickness dependent SOT characterization, we used the low-field harmonic Hall measurement technique[46-48].

The general linear-response expression for the out-of-plane ($\hat{z}$-directed) spin-current density generated by an in-plane electric field in a ferromagnet can be expressed as[11,13]

$$\boldsymbol{Q_z} = s_{//}[\boldsymbol{m} \cdot (\hat{z} \times \hat{E})]\boldsymbol{m} + s_{\perp}\boldsymbol{m} \times [(\hat{z} \times \hat{E}) \times \boldsymbol{m}] + s_{\perp}^{R}\boldsymbol{m} \times (\hat{z} \times \hat{E}), \quad [1]$$

Here, $\mathbf{m}$ is the unit magnetization vector and $\hat{E}$ is the unit vector along the applied in-plane electric field. $s_{//}$ parameterizes the component whose spin polarization is parallel to $\mathbf{m}$, while $s_{\perp}$, and $s_{\perp}^{R}$ describe the two symmetry-allowed transverse components (damping-like and field-like, respectively). These parameters represent only the symmetry weights of the respective contributions rather than physical response coefficients. Eq 1 exhausts all terms permitted by the symmetries of a ferromagnet for a spin current along $\hat{z}$.

We employed an in-plane, angle-dependent harmonic Hall measurement technique, which allows us to rule out spin currents generated by the anomalous Hall effect (AHE) and the planar Hall effect (PHE). These effects generate spin polarizations that are collinear with the magnetization; such polarizations undergo rapid spin dephasing within the ferromagnet and therefore do not contribute to a measurable torque[49]. Consequently, the detected second-harmonic Hall response is driven solely by spin-current components with a spin polarization transverse to $\mathbf{m}$. These components arise from phenomena such as the conventional spin Hall effect (CSHE) and the magnetization-dependent spin Hall effect (MSHE). For the specific experimental configuration considered here, where the equilibrium magnetization is aligned along $\hat{x}$. i.e., Eq 1 simplifies to $\boldsymbol{Q_z} = -s_{\perp}\hat{y} + s_{\perp}^{R}\hat{z}$ when $\mathbf{m}//\hat{x}$. (See Supplementary Information).

Notably, combining the macrospin analysis (See Supplementary Information for second harmonic Hall voltage from macrospin model) with the spin-current form from the simplified Eq 1, the 2ω transverse Hall resistance measured while rotating the sample in the *xy* plane by an angle $\varphi$ ($\varphi$ measured from +*x*, the current direction) can be written as

$$R_{xy}^{2\omega} = \left(R_{AHE}\frac{H_{AD,y}}{H_{ext}+H_k} + R_{ANE}\right)cos\varphi + (2R_{PHE}\frac{H_{AD,z}}{H_{ext}})cos2\varphi + (2R_{PHE}\frac{H_{FL,y+Oe}}{H_{ext}})cos2\varphi cos\varphi \ + R_{AHE}\frac{H_{FL,z+Oe}}{H_{ext}+H_k} + R_{PNE}sin2\varphi \quad [2]$$

Here, $R_{PHE}$ & $R_{AHE}$ are the planar and anomalous Hall resistances. The anomalous Nernst resistance ($R_{ANE}$) arises from an out of plane thermal gradient perpendicular to both the temperature gradient and the magnetization, while the planar Nernst contribution ($R_{PNE}$)[50] originate from in-plane thermal gradient along and perpendicular to the magnetization within the material's plane. $H_{AD}$ & $H_{FL}$ are the SOT generated anti-damping and field like effective fields with subscripts $y$ & $z$ denoting the polarization direction. $H_k$ is the effective perpendicular anisotropy field estimated through out of plane field sweep of the sample (see supplement). $H_{ext}$ is the external magnetic field applied during the rotation sweep and $H_{FL}$ includes Oersted field contribution, if any.

For compact fitting we define

$$R_{xy}^{2\omega} = R_{AD}^{y,2\omega}cos\varphi + R_{AD}^{z,2\omega}cos2\varphi + R_{FL+Oe}^{y,2\omega}cos2\varphi cos\varphi \ + R_{FL+Oe}^{z,2\omega} + R_{PNE}sin2\varphi \quad [3]$$

When we rotate the sample in the xy plane, with current along *x* the angular parts of $R_{xy}^{2\omega}$ that are symmetric under $\mathcal{M}_{yz}$ mirror operation arises from spin-orbit-torque terms with *y & z* -polarized spins (anti-damping and field-like, respectively), while the asymmetric part contains the thermoelectric (PNE) contributions. To isolate these effects experimentally, we symmetrize and antisymmetrize $V_{xy}^{2\omega}$about $\varphi = 180°$ to cleanly isolate the *y & z* -polarized SOT terms from the thermoelectric PNE signals. (See supplementary information for the detailed calculation)

Fig. 2(a) presents the typical second harmonic Hall response of the 14.4 nm sample for different values of $H_{ext}$. In Fig. 2(b), the raw data are decomposed into symmetric and antisymmetric components, which are then fitted using the respective terms of Eq 3. Since the torque related to the z-polarized spin current is negligible in our system (see Supplementary Information), our subsequent analysis focuses solely on the y-polarization. The field dependences of the fit coefficients associated with the y-polarized spin current were extracted using Eq. 3, as plotted in Fig. 2(c,d) for the 14.4 nm-thick sample.

The in-plane external field ($H_{ext}$) used during the rotation sweeps were chosen large enough to align $\boldsymbol{m} // H_{ext}$, yet modest enough to avoid measurable ordinary Nernst effect (ONE). Since $R_{PHE}$ is field-independent over the range studied and the SOT effective field ($H_{FL,y+Oe}$) is field-independent, the SOT portions of $\frac{R^{y,2\omega}_{FL+Oe}}{2R_{PHE}}$ should scale as $^1/_{H_{ext}}$(see Eq 2&3). Coefficient $\frac{R^{y,2\omega}_{FL+Oe}}{2R_{PHE}}$ indeed follows an inverse trend, consistent with a substantial $y$-polarized field-like torque $H_{FL,y+Oe}$ (and/or Oersted contribution) that is essentially field independent in the measurement window.

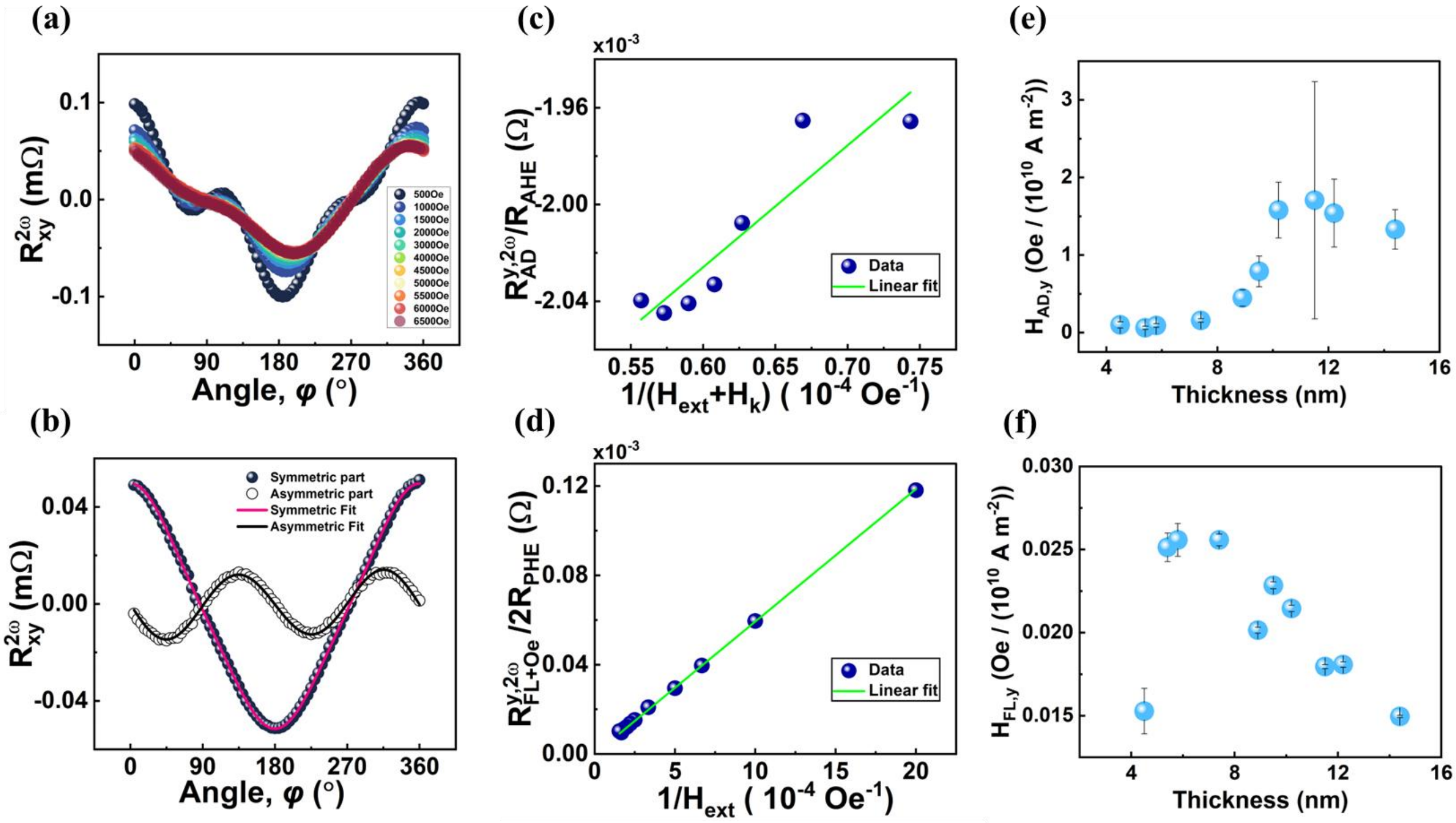

FIG. 2. (a) Angular dependence of the second harmonic Hall voltage $R^{2\omega}_{xy}$ as a function of φ for t =14.4nm under varying external magnetic fields ($H_{ext}$) measured using $J_{RMS}$ = $2.34 \times 10^{10}$ A/m$^2$. (b) Raw data separated into symmetric and asymmetric parts with fit. (c) & (d) External field dependence of the fit coefficients for NiFe of 14.4 nm thickness, extracted from the $R^{2\omega}_{xy}$ analysis. (e) Damping-like, and (f) field-like effective fields per current density as a function of NiFe thickness, for y-polarized spin currents. Errors are standard errors from the fitting.

As anticipated from $H_k \gg H_{ext}$, the coefficients $\frac{R^{y,2\omega}_{AD}}{R_{AHE}}$ show little dependence on $H_{ext}$ (Fig. 2(c)), remaining nearly constant within experimental field range. The residual offsets observed in $\frac{R^{y,2\omega}_{AD}}{R_{AHE}}$ are consistent with an anomalous Nernst contribution $R_{ANE}$, which arises from any heat generated in the system. Since the thermal conditions remain constant across the field range, the ANE-induced offset is also constant and can be extracted as the baseline. Notably, lower thickness samples show a larger positive offset in $\frac{R_{AD}^{y,2\omega}}{R_{AHE}}$, pointing to an enhanced anomalous Nernst contribution $R_{ANE}$ that grows with increasing resistivity (and thus Joule-heating-driven $\nabla_{Tz}$ ) at low NiFe thickness. We independently confirmed PNE by verifying the expected $J^2$ scaling of the corresponding 2ω signal (See Supplementary Sec.s9). The complete data for all the thicknesses is provided in Supplementary Sec.s8 more with a detailed explanation related to fitting and thermoelectric contribution. Using the relations in Eq 3 and thickness-specific calculations of $R_{AHE}$, $R_{PHE}$, & $H_k$(Supplementary information), we extract the SOT effective fields $H_{AD,y}$ and $H_{FL,y}$ per current density for each thickness as shown in Figure 2e & 2f respectively.

Fig. 2e & 2f reveals the existence of considerable SOT field in symmetric MgO/NiFe/MgO structure which varies with the thickness of NiFe. Although spin–orbit coupling in NiFe generates spin currents, the net torque for an ideal symmetric NiFe layer is expected to vanish

$$\tau_{\text{net}} = \frac{\hbar}{2e}[j_s(\text{ top }) - j_s(\text{ bottom })] = 0 \qquad [5]$$

When the two interfaces are in contact with dissimilar spin-sinks[33,51] or if there are spatial variations in the structural or magnetic properties within the ferromagnet[12,27], uncompensated spin current can be generated, resulting in the observed net self-torque.

Notably, in our MgO/NiFe/MgO trilayer structure, both the MgO/NiFe (top) and NiFe/MgO (bottom) interfaces were engineered to be identical which should lead to the cancellation of any such torque [52]. In order to check whether the slight thickness gradient from wedge deposition might break this symmetry, we fabricated a Hall bar oriented perpendicular to the wedge ensuring no thickness gradient exists along the current path. This device exhibits nearly identical spin-orbit-torque (SOT) efficiency (Supplementary Table s1), ruling out the wedge gradient as the dominant source of the self-torque.

Subsequently, we carried out XPS depth profiling studies on a sample with 12.2 nm thickness of NiFe starting from the top and going deeper into the MgO-NiFe-MgO-$SiO_2$ stack (for details, see supplement s6). The chosen levels represent different depths within the stack, with etch levels 3-15 representing the top MgO layer, the MgO/NiFe (top) interface, and the majority of the NiFe layer, while etch levels 27-39 represent the NiFe/MgO (bottom) interface, the bottom MgO layer, and the underlying $SiO_2$ substrate. Since the metal oxide peaks of Mg, Ni and Fe overlap, it is difficult to ascertain their individual contributions from the O1s peak alone (Fig. s6); therefore, studying the signatures in Ni and Fe provides more meaningful insight. The Ni *2p* core level spectra (Fig. 3a) shows the characteristic peak of Ni metal at ~ 852.6 eV until etch level 15 along with its satellite peaks, while additional peaks at ~ 853.7 eV corresponding to NiO start appearing from etch level 27. This implies that the bottom NiFe/MgO has some Ni oxidation while the top MgO/NiFe remains free of detectable Ni oxidation. A similar trend is also visible in Fe *2p* core level spectra (Fig. 3b) where etch level 15 shows the signature peaks of the metallic Fe (~706.7 eV) while at higher etch levels, we start obtaining peaks at higher binding energies corresponding to oxides of Fe ($Fe_2O_3$ ~709.4 eV).

This observation suggests that the oxidation-induced magnetic dead layer at the bottom interface breaks the structural inversion symmetry, creating the conditions necessary for net torque generation. Supporting this, SQUID magnetometry indicates a dead-layer thickness of ≈1.8 nm alongside a bulk-like saturation magnetization, $M_s = 772\ emu/cm^3$. A fit of the magnetic moment/area versus NiFe thickness is given in Fig. 3c. We therefore attribute the observed self-torque to the interfacial asymmetry arising from the magnetic dead layer.

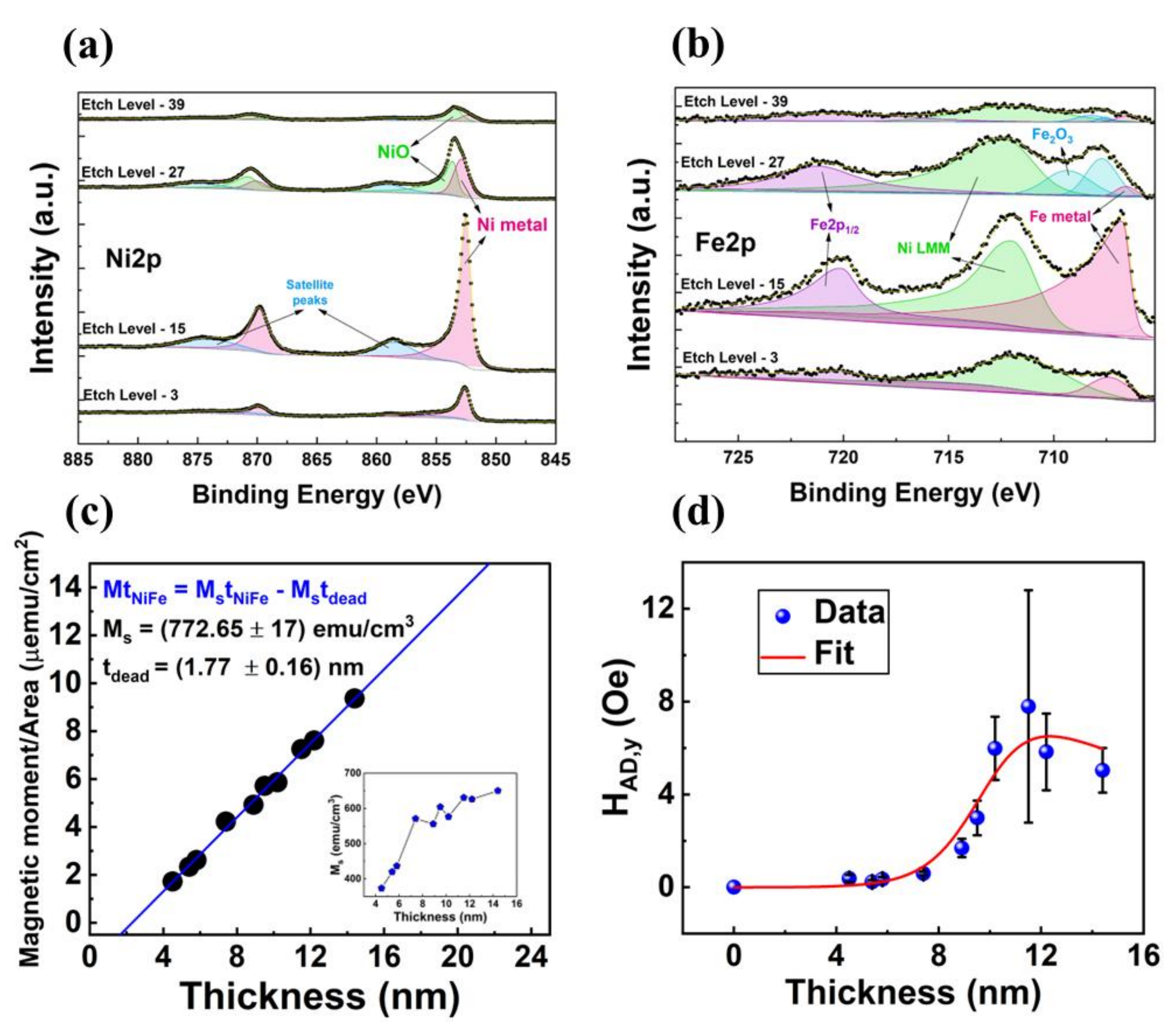

FIG. 3. XPS core level spectra analysis of (a) Ni2p and (b) Fe2p using depth profiling. The presence of oxides in etch level 27 depicts the presence of the dead layer in the NiFe/MgO (bottom) interface. (c) Magnetic moment/Area vs thickness of NiFe measured using SQUID magnetometry. (d) Spin orbit torque effective field vs thickness of NiFe fitted using Eq 6.

As shown in Fig. 2e & 2f, the SOT effective field arising from the y-polarized spin current initially increases with NiFe thickness, reaches a maximum, and then gradually decreases, whereas the field-like (FL) torque remains non-negligible throughout. To interpret this behavior, we adapt the Kim–Lee drift-diffusion formalism[1,53] by incorporating different spin-current absorption coefficients at the top and bottom interfaces, reflecting the influence of the dead layer. According to their model, a ferromagnet can generate a transverse spin current that exerts a “self-torque” on its own magnetization. Kim and Lee showed that a damping-like spin Hall current generated within the ferromagnet can precess while diffusing through the magnetic bulk, that accumulates at the interface and increases with thickness before eventually saturating. This mechanism provides direct evidence for substantial self-torque in single-layer NiFe systems.

To quantitatively capture this behavior, we employ the full expression (see supplement):

$$H_{\mathrm{AD}}(t) = \frac{C_{\mathrm{self}}}{t_{\mathrm{eff}}} \times \mathrm{Re}\left[\frac{(g_{\mathrm{top}} - g_{\mathrm{bot}})\lambda_F \tanh(t_{\mathrm{eff}}/2\lambda_F)}{(g_{\mathrm{top}} + g_{\mathrm{bot}})\lambda_F \coth(t_{\mathrm{eff}}/\lambda_F) + K}\right] \quad [6]$$

This formulation simultaneously accounts for the saturation and 1/t dependence of the damping-like (DL) effective field. Physically, two competing effects coexist within a single ferromagnetic layer. The first contribution leads to saturation beyond a certain thickness, where the ferromagnet itself acts as a spin-current source. The second one introduces a 1/t dependence, reflecting the dilution of spin angular momentum in thicker films, which reduces the torque per unit magnetization. The observed thickness dependence is inconsistent with a purely interfacial Rashba mechanism, which would yield a decreasing trend with thickness. Fitting the experimental DL torque data with this model yields an estimated magnetically dead-layer thickness of ~1.2 nm (Figure 3d), which is in close agreement with the independently measured value of ~1.8 nm. Meanwhile, the FL torque exhibits non-negligible values across the thickness range, consistent with the Kim–Lee model, where spin precession within the ferromagnet generates transverse field-like torque components even in the absence of interfacial asymmetry.

This analysis underscores how even subtle interface asymmetries can enable significant self-induced SOTs in a nominally symmetric NiFe, effectively bridging theory with our experimental observation. Previous studies on interfacial engineering[54] have shown that structural and chemical asymmetry at interfaces can tune the relative strengths of the observed torques; however, such approaches typically require deliberate modifications to tailor spin transparency and interfacial spin–orbit coupling often accompanied by increased current shunting. Thus, the naturally occurring dead layer in our system provides an intrinsic source of asymmetry that can produce comparable modifications to the torque balance. It is intuitive to compare the self-induced SOT observed here with conventional HM/FM bilayers. In Pt/FM bilayers, the damping-like SOT arises primarily from the Pt spin Hall effect[55-58]. In contrast, our MgO/NiFe/MgO system, which contains no heavy metal, exhibits a damping-like effective field without any deliberate spin-current source — it is generated entirely by the intrinsic SHE of NiFe itself, made observable by the dead-layer-induced symmetry breaking. Moreover, while the spin Hall effect can, in principle, contain both intrinsic and extrinsic (skew-scattering and side jump) contributions, the latter are expected to be negligible in concentrated alloys such as $Ni_{80}Fe_{20}$. Extrinsic mechanisms are most significant in the dilute impurity limit, whereas in concentrated transition-metal alloys the intrinsic band structure contribution is known to dominate, as established by both theoretical[59] and experimental[60] studies. Consequently, the damping like torque observed in our system is most naturally attributed to the intrinsic spin Hall effect of NiFe.

To support the experimental findings, we performed density functional theory (DFT) calculations on the (111) surface of an fcc $Ni_3Fe$ alloy (approximate composition $Ni_{80}Fe_{20}$) (see supplement for the XRD) to estimate the intrinsic spin Hall conductivity (SHC). The geometrical model is displayed in Fig. 4a. The (111) plane is the close-packed plane in FCC structures, with atoms arranged in a hexagonal lattice. In (111) surface, each layer consists of 75% Ni and 25% Fe atoms in an ordered pattern. The stacking sequence along [111] directions is ABC as indicated in Fig. 4a, but all layers have identical composition. The electronic band structure calculation reveals a metallic surface with multiple bands intersecting the Fermi level, corroborated by the Fermi surface plot in Fig. 4(b-c). The (111) surface exhibits an intrinsic SHC ($\sigma_{xy}^{z}$) magnitude of ~ $3\times10^2$ ($\hbar$/e) S/cm, which is about an order of magnitude lower than that of Pt (~$2\times10^3$ ($\hbar$/e) S/cm)[61] as shown in Figure 4d but comparable to that of CdTe and ZnTe (110) facets[62] and $MnPd_3$ (114) film[16]. Here $\sigma_{\alpha\beta}^{\gamma}$ denotes $\alpha$:direction of spin current, $\beta$:direction of applied electric field/current, $\gamma$:spin polarization direction. Nevertheless, the SHC plot indicates that NiFe intrinsically supports sizable spin currents, consistent with experimentally measured values at room temperature. To further understand the experimentally observed thickness dependence, the electronic band structure and intrinsic SHC are calculated and presented in Fig. s15. As the number of layers increases, more bands cross the Fermi level, in agreement with the Fermi surface plot. Consequently, the intrinsic SHC also increases with increasing layers, as shown in Fig. 4e.

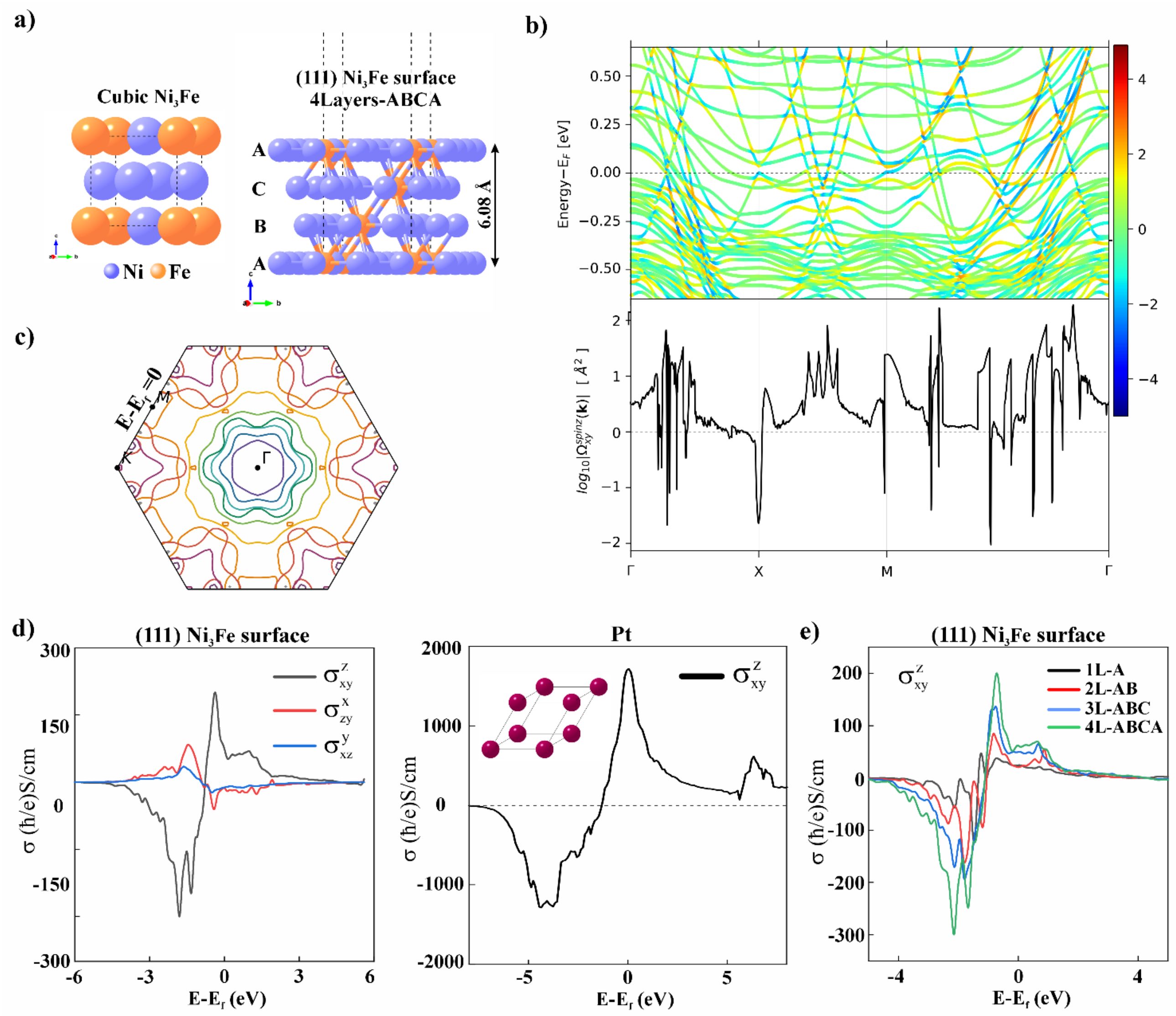

FIG. 4. DFT Analysis (a) Geometrical view of fcc $Ni_3Fe$ and optimized (111) surface of $Ni_3Fe$, (b) spin Hall conductivity (SHC) projected band structure along with k-point resolved SHC, (c) Fermi surface, (d) direction-dependent intrinsic SHC of $Ni_3Fe$ surface, and that of Pt for comparison, (e) comparative plot of layer-dependent intrinsic SHC.

To investigate the SHC of the heterointerface corresponding to the experimentally studied system, a model heterointerface is constructed comprising the MgO (111) and $Ni_3Fe$ (111) surfaces. The optimized geometry of the MgO-$Ni_3Fe$ (4L) heterointerface is presented in Fig. 5(a). The optimized structure reveals the formation of interfacial bonds between the O atoms of the MgO layer and the $Ni_3Fe$ surface, indicating strong interfacial coupling. The MgO-projected band structure shows that the electronic states originating from MgO are far away from the Fermi level, whereas the states in the vicinity of the Fermi level are predominantly contributed by the $Ni_3Fe$ layer. The SHC of the MgO/$Ni_3Fe$ heterointerfaces with varying $Ni_3Fe$ thickness is presented in Fig. 5(c). In these calculations, the thickness of the O-terminated MgO layer was kept fixed, while the number of $Ni_3Fe$ layers was varied. The results show that the SHC of the heterointerfaces is slightly reduced compared to the corresponding pristine Ni3Fe slabs. However, the overall thickness-dependent trend remains essentially unchanged, with the SHC increasing consistently as the $Ni_3Fe$ layer thickness increases. Furthermore, the DFT calculations reveal that MgO contributes negligibly to the SHC, as illustrated in Fig. 5(c), owing to its electronic structure being dominated by s and p orbitals, which exhibit weak spin-orbit coupling.

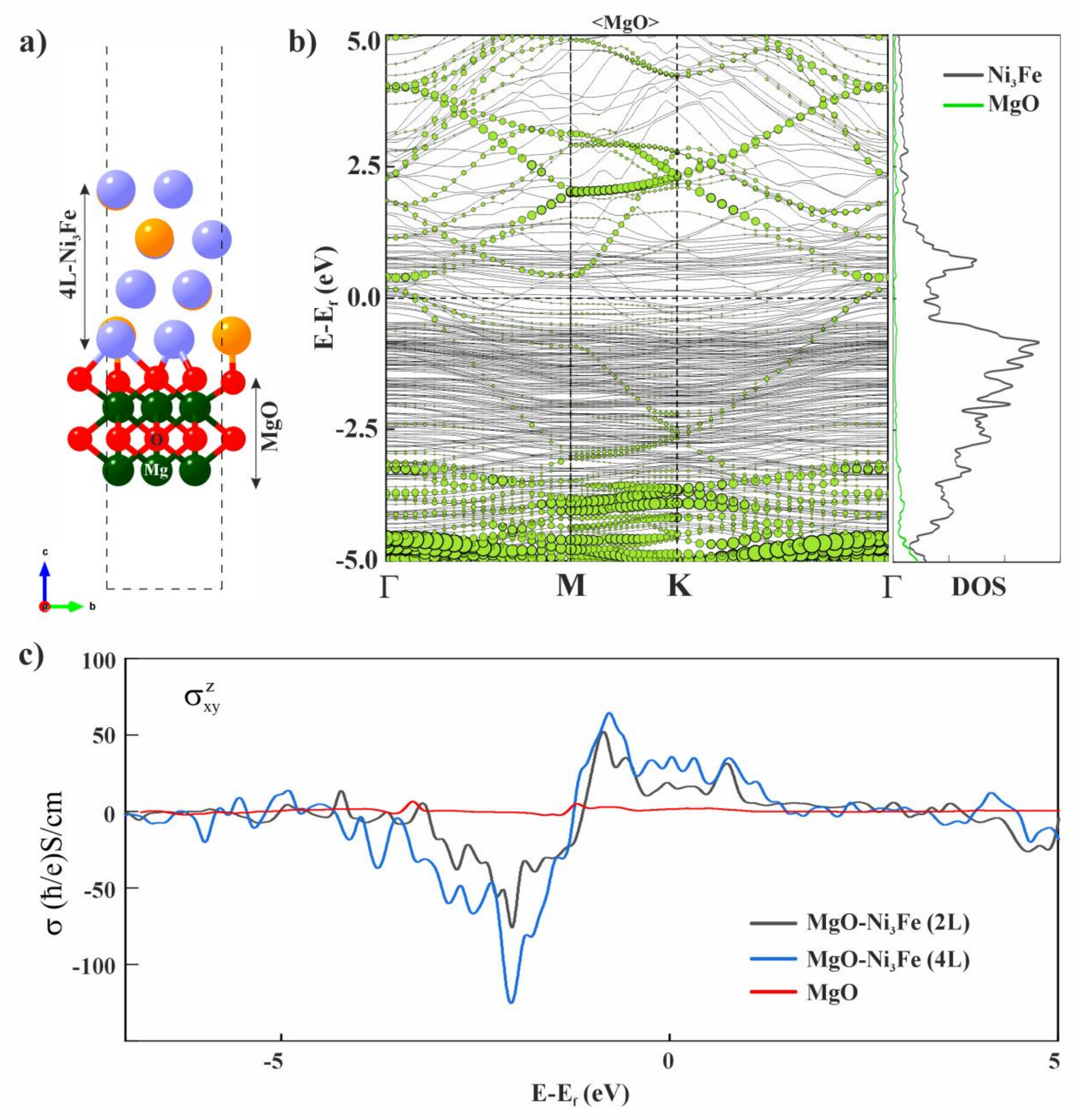

FIG. 5. (a) Optimized atomic structure of $MgO-Ni_3Fe$ (4L) heterointerface, (b) Electronic band structure calculated including spin-orbit coupling (SOC), with the MgO contribution projected onto the bands, (c) SHC

of the MgO-$Ni_3Fe$ heterointerface with two-layers and four-layer $Ni_3Fe$, together with that of pristine MgO for comparison.

Notably, although an MgO layer is present both above and below, the (111) $Ni_3Fe$ surface terminations at the top and bottom interfaces may not be equivalent[63]. Due to the ABC-type stacking sequence, the bonding configurations and electronic hybridization at the top MgO/$Ni_3Fe$ interface can be inherently different from those at the bottom interface. Although MgO may not generate a spin Hall effect, it hybridizes with $Ni_3Fe$ orbitals at the interface, modifying SOC and orbital moments. Such interfacial asymmetry is further amplified by the presence of a magnetic dead layer at the bottom NiFe/MgO interface. However, confirming the exact atomic terminations of the (111) surface requires precise structural characterization, which lies beyond the scope of this work. This experimental work also indicate that the spin transparency differs at the top and bottom interfaces due to the dead layer, ultimately gives rise to a sizeable self-torque—even in a nominally symmetric MgO/NiFe/MgO trilayer.

## CONCLUSION

In conclusion, we demonstrate that a nominally symmetric MgO/NiFe/MgO trilayer produces sizable spin–orbit torques despite the absence of heavy metals and without engineered structural asymmetry. XPS and SQUID measurements reveal a naturally occurring magnetic dead layer at the bottom interface, which provides the symmetry breaking required to generate an uncompensated spin current. The full thickness dependence of the torque is quantitatively captured by the Kim–Lee drift–diffusion formalism, and density-functional calculations confirm that NiFe possesses sufficient intrinsic spin Hall conductivity to support the effect. These results establish a previously unrecognized source of self-induced torque in common ferromagnets and demonstrate that interfacial dead layers—typically considered parasitic—can act as functional spin-conversion elements. A natural extension of this work would be to systematically vary the dead-layer nature—for example, by oxidizing MgO seed layers under controlled $O_2$ partial pressures or in presence of reactive oxygen prior to NiFe deposition—thereby enabling direct, tunable control over the interfacial symmetry breaking and the resulting torque magnitude. In short, the very interface defects long dismissed as detrimental may hold the key to a new generation of efficient, heavy-metal-free spintronic devices.

## ACKNOWLEDGEMENT

ARK acknowledges the financial assistance from DST-INSPIRE. PJ acknowledges the support provided by the U.S. Department of Energy, Office of Basic Energy Sciences, Division of Materials Sciences and Engineering under Award No. DE-FG02-96ER45579. Resources of the National Energy Research Scientific Computing (NERSC) Center is supported by the Office of Science of the U.S. Department of Energy under Contract No. DE-AC02-05CH11231 is also acknowledged. The authors extend their acknowledgment to the High-Performance Research Computing (HPRC) core facility at Virginia Commonwealth University for providing supercomputing resources. MKM acknowledges the ANRF Ramanujan Faculty fellowship under Award No. RJF/2025/000601 for financial support. A.A. acknowledges support through project 3D-Sky (Grant No. GAP-101108063), part of the Marie Skłodowska- Curie Actions under the Horizon Europe Program of the EU. DR acknowledge the financial support from the Department of Atomic Energy (DAE) under project no. 58/20/10/2020-BRNS/37125 & Anusandhan National Research Foundation (ANRF) under project no. ANRF/ARG/2025/009882/PS.

## DATA AVAILABILITY

The data that support the findings of this study are available from the corresponding author upon reasonable request.